\documentclass[%
 amsmath,amssymb,
preprint,%
 reprint,%
]{revtex4-1}

\usepackage{graphicx}
\usepackage{dcolumn}
\usepackage{bm}

\begin{document}

\preprint{AIP}

\title{Vibration Powered Radiation of Quaking Magnetar}

\author{S.  Bastrukov}

\affiliation{State Key Laboratory of Nuclear Physics and Technology,\\
 School of Physics, Peking University, 100871 Beijing, China}

 \author{I. Molodtsova}
\affiliation{Joint Institute for Nuclear Research, 141980 Dubna, Russia}

\author{J. W. Yu}
\affiliation{State Key Laboratory of Nuclear Physics and Technology,\\
 School of Physics, Peking University, 100871 Beijing, China}

\author{R. X. Xu}
\affiliation{State Key Laboratory of Nuclear Physics and Technology,\\
 School of Physics, Peking University, 100871 Beijing, China}

\date{\today}

\begin{abstract}
In  juxtaposition with the standard model of rotation powered pulsar, the 
model of vibration powered magnetar undergoing quake-induced 
torsional Alfv\'en vibrations in its own ultra strong magnetic field experiencing decay is considered.  The presented line of argument suggests  that gradual decrease of  
frequencies (lengthening of periods) of long-periodic pulsed radiation 
detected from a set of X-ray sources can be attributed to magnetic-field-decay induced energy conversion from seismic vibrations to magneto-dipole radiation of quaking magnetar.
\end{abstract}



\maketitle

\section{Introduction}

 There is a common recognition today that the standard (lighthouse) model of 
 inclined rotator, lying at the base of current understanding of radio pulsars, faces 
 serious difficulties in explaining  the long-periodic ($2<P<12$ s) pulsed radiation of 
 soft gamma repeaters (SGRs) and anamalous $X$-ray pulsars (AXPs). The 
 persistent $X$-ray luminosity of AXP/SGR sources      
 ($10^{34}<L_X<10^{36}$ erg  s$^{-1}$) is  appreciably (10-100 times) larger 
 than expected from a neutron star deriving radiation power from the 
 energy of rotation with frequency of detected pulses. 
 Such an understanding has come soon after the detection on March 5, 1979 
 of the first  0.2 second long gamma burst \cite{M-79} which  
 followed by a 200-seconds emission that showed a clear 8-second pulsation period  
 \cite{B-79} and association of this event to a supernova remnant known as N49 in  
 the  Large Magellanic Cloud \cite{CL-80}. This object
 is very young (only a few thousand years old) but period of pulsating emission is 
 typical of a much older neutron star. In works \cite{RN-80,RST-80}
 it has been proposed that discovered object, today designated SGR 0526-66, 
 is a vibrating neutron star, that is, the detected for the first time long-periodic pulses 
 owe their origin to the neutron star vibrations, rather than rotation as is the case   
 radio pulsars. During the following decades, the study of these objects has been 
 guided by idea \cite{DT-92,P-92} that electromagnetic activity 
 of magnetars both AXP's and SGR's, is primarily determined by decay of ultra strong 
 magnetic field ($10^{14}<B<10^{16}$ G) and that a highly intensive gamma 
 bursts are manifestation of magnetar quakes \cite{BL-89,CE-96,DC-97}.
 
 In this work we investigate in some details the model of vibration powered magnetar  
 which is in line with the current treatment 
 of quasi-periodic oscillations of outburst luminosity of soft gamma repeaters as 
 being produced by Lorentz-force-driven torsional seismic vibrations  
 triggered by quake. As an extension of this point of view, in this paper 
 we focus on impact of the magnetic field decay on Alfv\'en vibrations  
 and magneto-dipole radiation generated by such vibrations. 
 Before so doing, it seems appropriate to  
 recall a  seminal paper of Woltjer \cite{W-64} who was first to observe 
 that magnetic-flux-conserving core-collapse supernova can produce 
 a neutron star with the above magnetic field intensity of typical magnetar. 
 Based on this observation, Hoyle, Narlikar and Wheeler \cite{HNW-64} proposed that a strongly magnetized 
 neutron star can generate magneto-dipole radiation powered by energy of 
 hydromagnetic, Alfv\'en, vibrations stored in the star after its birth in supernova 
 event (see, also, \cite{P-08}). Some peculiarities of this mechanism of vibration 
 powered radiation has been scrutinized in our recent work \cite{B-11}, 
 devoted to radiative activity of pulsating magnetic white dwarfs, in which it was 
 found that necessary condition for the energy conversion from Alfv\'en vibrations 
 into electromagnetic radiation is the decay of magnetic field. As was stressed,
 the magnetic field decay is one of the most conspicuous 
 features distinguishing magnetars from normal rotation powered 
 pulsars. It seems not implausible, therefore, to expect that at least some of currently 
 monitoring  AXP/SGR - like sources are magnatars  
 deriving power of pulsating magneto-dipole radiation from 
 energy of Alfv\'enic vibrations of highly conducting matter 
 in the ultra strong magnetic field experiencing decay. 
 
 In approaching Alfv\'en vibrations of neutron star in its own 
 time-evolving magnetic field, we rely on the results of recent investigations 
 \cite{B-WS,B-09a,B-09b,B-10c} of both even parity poloidal 
 and odd parity toroidal (according to Chandrasekhar terminology \cite{Ch-56})  node-free Alfv\'en vibrations of magnetars in constant-in-time magnetic field.  The  extensive review of earlier investigations of standing-wave regime 
 of Alfv\'enic stellar vibrations can be found in \cite{LW-58}. 
 The spectral formula for discrete frequencies of both poloidal and toroidal $a$-modes   
 in a neutron star with mass $M$, radius $R$ and magnetic field of typical magnetar, $B_{14}=B/10^{14}\,$ G, reads 
    \begin{eqnarray}
 \label{e1.1}
 &&\omega_\ell=\omega_A\,s_\ell,\,\,\omega_A=\frac{v_A}{R}=B\sqrt{\frac{R}{3M}},\,\,M=\frac{4\pi}{3}\rho R^3,\\
  \label{e1.2}
 &&\nu_{\rm A}=\frac{\omega_A}{2\pi}=0.2055 B_{14} R^{1/2}_{6}, (M/M_{\odot})^{-1/2},\,\, {\rm Hz}.
\end{eqnarray}
where numerical factor $s_\ell$ is unique to each specific shape of 
magnetic field frozen in the neutron star of one and the same mass  $M$ and radius $R$.

\section{Alfv\'en vibrations of magnetar in 
time-evolving magnetic field}

  In above cited work it was shown that Lorentz-force-driven 
  shear node-free vibrations of magnetar in its own magnetic field field ${\bf B}$  
  can be properly described in terms of material displacements ${\bf u}$ 
  obeying equation of magneto-solid-mechanics
 \begin{eqnarray}
  \label{e2.1}
 && \rho\, {\ddot {\bf u}}({\bf r},t)=\frac{1}{4\pi}
 [\nabla\times[\nabla\times [{\bf u}({\bf r},t)\times {\bf B}({\bf r},t)]]]\times {\bf B}
 ({\bf r},t)\\
 \label{e2.2}
 && {\dot {\bf u}}({\bf 
   r},t)=[\mbox{\boldmath $\omega$}({\bf r},t)\times {\bf r}],\quad 
   \mbox{\boldmath $\omega$}
  ({\bf r},t)=A_t[\nabla\chi(r)]\,{\dot\alpha}(t).
  \end{eqnarray}
  The field ${\dot {\bf u}}({\bf r},t)$ is identical to that for torsion node-free 
  vibrations restored by  Hooke's force of elastic stresses  \cite{B-10c,B-07}
  with  $\chi({\bf r})=A_\ell\, f_\ell(r)\,P_\ell(\cos\theta)$ where 
  $f_\ell(r)$ is the nodeless function of distance from the star center and 
   $P_\ell(\cos\theta)$ is Legendre polynomial of degree $\ell$ specifying the overtone 
   of toroidal mode. In (\ref{e2.2}), the amplitude ${\alpha}(t)$ is the basic 
   dynamical variable describing time evolution of vibrations which is different for 
   each specific overtone; in what follows we confine our analysis to solely 
   one quadrupole overtone.
   The central to the subject of our study is the following representation 
   of the time-evolving internal magnetic field 
  \begin{eqnarray}
  \label{e2.3}
 && {\bf B}({\bf r},t)=B(t)\,{\bf b}({\bf r}), 
 \end{eqnarray}
  where $B(t)$ is the time-dependent intensity 
  and ${\bf b}({\bf r})$ is dimensionless vector-function of the field distribution 
  over the star volume.  Scalar product of (\ref{e1.1}) with the separable form 
 of material displacements 
 \begin{eqnarray}
  \label{e2.4}
 && {\bf u}({\bf r},t)={\bf a}({\bf r})\,\alpha(t)
 \end{eqnarray}
 followed by integration over 
 the star volume leads to equation for amplitude $\alpha(t)$ 
 having the form of equation of oscillator with time-depended spring constant  
 \begin{eqnarray}
  \label{e2.5}
 && {\cal M}{\ddot \alpha}(t)+{\cal K}(t)\alpha(t)=0.
 \end{eqnarray}
   The total vibration energy and frequency are 
   given by
    \begin{eqnarray}
 \label{e2.6}
  && E_A(t)=\frac{{\cal M}{\dot \alpha}^2(t)}{2}+\frac{{\cal K}(B(t))\alpha^2(t)}
  {2},\\
  && \omega(t)=\sqrt{\frac{{\cal K}(t)}{\cal M}}=B(t)\kappa,\quad\kappa=\sqrt{\frac{R}{3M}}\,s.
  \end{eqnarray}
  It follows  
  \begin{eqnarray}
  \label{e3.2}
&& \frac{dE_A(t)}{dt}=
  \frac{\alpha^2(t)}{2}\frac{d{\cal K}(B)}{dB}\,\frac{dB(t)}{dt}.
  \end{eqnarray}
   This shows that the variation in time of magnetic field intensity in quaking 
   magnetar causes 
   variation in the vibration energy. In next section we focus on 
   conversion of energy of  Lorentz-force-driven
   seismic vibrations of magnetar into the energy of magneto-dipole radiation.

 \section{Vibration powered radiation of quaking magnetar}

    The point of departure in the study of vibration-energy powered
    magneto-dipole emission of the star (whose radiation power, ${\cal P}$, is given by Larmor's
    formula) is the equation
 \begin{eqnarray}
 \label{e3.4}
  && \frac{dE_A(t)}{dt}=-{\cal P}(t),\quad {\cal P}(t)=\frac{2}{3c^3}\delta {\ddot {\mbox{\boldmath
  $\mu$}}}^2(t).
  \end{eqnarray}
   We consider a model of quaking neutron star whose torsional magneto-mechanical
  oscillations are accompanied by fluctuations of total magnetic moment preserving
  its initial (in seismically quiescent state) direction: $\mbox{\boldmath
  $\mu$}=\mu\,{\bf n}={\rm constant}$.
  The total magnetic dipole moment should execute oscillations with frequency
  $\omega(t)$ equal to that for magneto-mechanical vibrations of  stellar matter which
  are described by equation for $\alpha(t)$.
  This means that $\delta {\mbox{\boldmath $\mu$}}(t)$ and  $\alpha(t)$
  must obey equations of similar form, namely
  \begin{eqnarray}
   \label{e3.5}
  && \delta {\ddot {\mbox{\boldmath $\mu$}}}(t)+\omega^2(t)
  \delta {\mbox{\boldmath $\mu$}}(t)=0,\\
   \label{e3.6}
  && {\ddot \alpha}(t)+\omega^2(t){\alpha}(t)=0,\quad \omega^2(t)=B^2(t)
  {\kappa}^2.
  \end{eqnarray}
  It is easy to see that equations (\ref{e3.5}) and  (\ref{e3.6}) can be
  reconciled if
  \begin{eqnarray}
   \label{e3.7}
  \delta \mbox{\boldmath $\mu$}(t)=\mbox{\boldmath $\mu$}\,\alpha(t).
  \end{eqnarray}
  Given this, we arrive at the following law of magnetic field decay 
   \begin{eqnarray}
  \label{e3.10}
  && \frac{dB(t)}{dt}=-\gamma\,B^3(t),\quad
  \gamma=\frac{2\mu^2\kappa^2}{3{\cal M}c^3}={\rm
  constant},\\
   \label{e3.11}
  && B(t)=\frac{B(0)}{\sqrt{1+t/\tau}},\quad \tau^{-1}=2\gamma B^2(0).
  \end{eqnarray}
The last equation shows the lifetime of quake-induced vibrations in question substantially depends 
upon the intensity of initial (before quake) magnetic field $B(0)$: the larger $B(0)$
the shorter $\tau$. For neutron stars with one and the same mass $M=1.4M_\odot$
and radius $R=15$ km, and magnetic field of typical pulsar $B(0)=10^{12}$ G, we 
obtain $\tau\sim 3\times 10^{7}$ years, whereas for  
magnetar with $B(0)=10^{14}$ G, $\tau\sim 7\times 10^{3}$ years.    
 
The equation for vibration amplitude   
\begin{eqnarray}
\label{ee1}
 &&{\ddot \alpha}(t)+\omega^2(t)\alpha(t)=0,\\
 \label{ee1a}
 && \omega^2(t)=\frac{\omega^2(0)}{1+t/\tau},\,\,\,\omega(0)=\omega_A\kappa.
  \end{eqnarray}
  with help of substitution $s=1+t/\tau$ is transformed to the equation 
\begin{eqnarray}
\label{ee8}
 s\alpha''(s)+\beta^2\alpha(s)=0,\quad \beta^2=\omega^2(0)\tau^2={\rm const}.
  \end{eqnarray}
  permitting general solution \cite{PZ-04}. The solution of this equation, obeying two 
  conditions $\alpha(t=0)=\alpha_0$ and $\alpha(t=\tau)=0$, can be represented 
  in the form 
   \begin{eqnarray}
\nonumber
 &&\alpha(t)=C\,s^{1/2}\{J_1(z(t))-\eta\,Y_1
 (z(t))\},\,\,z(t)=2\beta\,s^{1/2}(t)\\
 && \eta=\frac{J_1(z(\tau))}{Y_1(z(\tau))},\quad C=\alpha_0[J_1(z(0))-\eta\,Y_1(z(0)]^{-1}.
  \end{eqnarray}
 where $J_1(z)$ and $Y_1(z)$ are Bessel functions \cite{AS-72}
 and \begin{eqnarray}
\label{ee12a}
 &&\alpha_0^2=\frac{2{\bar E_A(0)}}{M\omega^2(0)}=\frac{2{\bar E_A(0)}}{K(0)},\quad
 \omega^2(0)=\frac{K(0)}{M}.
  \end{eqnarray}
 Here by ${\bar E}_A(0)$ is understood the average energy stored
 in torsional Alfv\'en vibrations of magnetar. If all the detected energy 
 $E_{\rm burst}$ of $X$-ray outburst goes 
  in the quake-induced vibrations, $E_{\rm burst}=E_A$, then the initial 
  amplitude $\alpha_0$ is determined unambiguously. 
  The impact of magnetic field decay on frequency and amplitude 
of torsional Alfv\'en vibrations in quadrupole overtone is illustrated
in Fig.1, where we plot $\alpha(t)$ with pointed out parameters $\beta$ and $\eta$. 
 \begin{figure}
 \centering
 \includegraphics[scale=0.4]{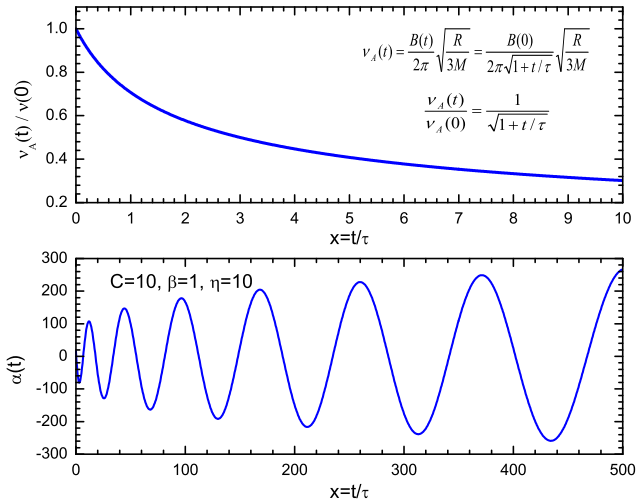}
 \caption{(Color online) The figure illustrates the effect of magnetic field decay on 
 the vibration frequency and amplitude of quadrupole toroidal $a$-mode presented 
 as functions of $x=t/\tau$.}
 \end{figure}
The magnetic-field-decay induced lengthening of period of pulsating radiation  
(equal to period of vibrations) is described by  
   \begin{eqnarray}
 \nonumber
 && P(t)=P(0)\,[1+(t/\tau)]^{1/2},\,\,
 {\dot P}(t)=\frac{1}{2\tau}\frac{P(0)}{[1+(t/\tau)]^{1/2}},\\
 && \tau=\frac{P^2(0)}{2P(t)\dot P(t)},\quad P(0)=\frac{2\pi}{\kappa\,B(0)}
 \label{e3.14}
 \end{eqnarray}
 On comparing $\tau$ given by equations (\ref{e3.11}) and (\ref{e3.14}), one 
 finds that interrelation between equilibrium equilibrium value of the total 
 magnetic moment $\mu$ of a neutron star of mass $M=1.4\,M_\odot$ and radius $R=10$ km vibrating in   
 quadrupole overtone of toroidal $a$-mode is given by 
  \begin{eqnarray}
     \label{e2.9}
&&\mu=A(M,R) \sqrt{P(t){\dot P}(t)},\,\,A=\sqrt{\frac{3{\cal M}c^3}{8\pi^2}},\\
\label{e2.10}
&& \mu=3.8 \times 10^{37}\,\sqrt{P(t)\,{\dot P}(t)},\,\,{\rm G\,cm^3}.
\end{eqnarray}
For a sake of comparison, 
in the considered model of vibration powered radiation, the equation
of magnetic field evolution is obtained in similar fashion
  as that for the angular velocity $\Omega(t)$ does in
  the standard model of rotation powered neutron star which rests on equations
  \begin{eqnarray}
  && \frac{dE_R}{dt}=-\frac{2}{3c^3}\delta {\ddot {\mbox{\boldmath
  $\mu$}}}^2(t),\\
  && E_R(t)=\frac{1}{2}I\,{\Omega}^2(t), \quad I=\frac{2}{5}MR^2\\
  && \delta {\ddot {\mbox{\boldmath
  $\mu$}}}(t)=[\mbox{\boldmath ${\Omega}$}(t)\times [\mbox{\boldmath ${\Omega}$}(t)\times {\mbox{\boldmath
  $\mu$}}]],\\ 
  && {\mbox{\boldmath
  $\mu$}}=\mu {\bf n},\, \mu=\frac{1}{2}BR^3
 \end{eqnarray}
which lead to 
\begin{eqnarray}
  &&{\dot \Omega}(t)=-K\Omega^3(t),\quad K=\frac{2\mu_{\perp}^2}{3Ic^3},\quad
  \mu_{\perp}=\mu\,\sin\theta,\\
\label{e2.6a}
&& \Omega(t)=\frac{\Omega(0)}{\sqrt{1+t/\tau}},\quad \tau^{-1}=2\,K\,\Omega^2(0).
 \end{eqnarray}  
  where $\theta$ is angle of inclination of $\mbox{\boldmath  $\mu$}$  to $\mbox{\boldmath  ${\Omega}$}(t)$. The time evolution of $P(t)$, $\dot P(t)$ and expression for $\tau$ are too described by equations (\ref{e3.14}). 
It is these equations which lead to widely used exact analytic estimate of magnetic field on the neutron star pole:
  $B=[3Ic^3/(2\pi^2\,R^6)]^{1/2}\sqrt{P(t)\,\dot P(t)}$. For a neutron star 
  of mass $M=M_\odot$, and radius $R=13$ km, one has $B=3.2\,10^{19} 
  \sqrt{P(t)\,\dot P(t)}$ G.
 Thus, the substantial physical difference between vibration and rotation 
 powered neutron star models is that  in the former model the elongation of pulse period is attributed to magnetic field decay, whereas in the latter one the period lengthening 
  is ascribed to the slow down of 
  rotation \cite{MT-77,LK-04}.

\section{Summary}
  The last  two decades have seen a growing understanding that magnetic 
  field decay is the key property distinguishing magnetars from pulsars.
   The magnetic fields frozen-in the immobile matter of pulsars (most of which are 
   fairly stable to starquakes)  operate like a passive (unaltered in time) promoter of 
      their persistent radiation along the axis of dipole magnetic moment of 
      the star inclined to the rotation axis. The presented treatment  
      of spin-down effect emphasizes kinematic nature of time evolution 
      of pulsar magnetic moment as brought about by rapid rotation with gradually 
      decreasing  frequency and showing that rotation does not affect intensity of 
      its internal magnetic field. 
      Unlike pulsars, magnetars (SGR-like sources) are  
      isolated quaking neutron stars whose seismic and radiative activity 
      is thought to be dominated by decay of magnetic field. Perhaps the most striking 
      manifest of seismic vibrations of magnetar are quasi-periodic oscillations 
      of outburst luminosity rapidly decreasing (in several hundred seconds) 
      from about $L_X\sim 10^{44}$ erg  s$^{-1}$ in 
      giant flare to about $L_X\sim10^{34}$ erg  s$^{-1}$ in quiescent regime 
      of long-periodic X-ray emission.  As was emphasized, 
      this lowest value of luminosity is well above of the rate of energy of 
      rotation with period equal to that of detected emission.   
      With all above in mind, we have set up  a model 
      of vibration powered radiation of quaking magnetar in which 
      the key role is attributed to magnetic field decay.
      The most striking outcome of presented line of argument 
      is gradual decrease of frequencies (lengthening of periods) of  
      magneto-dipole radiation (pulsating with the frequency 
      of toroidal $a$-mode of seismic vibrations) owing its origin to magnetic field 
      decay. 
      Remarkably that this 
      prediction of 
      the model is consistent with data on gradual decrease of frequency of pulsed $X$-
      ray emission 
      of such magnetars as 1E 2259.1+586 \cite{GKW-04} and XTE J1810-197    
      \cite{Ib-04}. It is hoped, therefore, that presented here theoretical 
      results can be efficiently utilized as a guide in observational quest for vibration powered 
      neutron stars.

\section*{Acknowledgment}

    This work is supported by the National Natural Science Foundation of China
(Grant Nos. 10935001, 10973002), the National Basic Research Program of
China (Grant No. 2009CB824800), and the John Templeton Foundation.

\end{document}